\documentclass[10pt,conference]{IEEEtran}
\usepackage{graphicx}
\usepackage{amsmath}
\usepackage{amssymb}
\usepackage[colorlinks]{hyperref}
\usepackage{theorem}

\theoremstyle{plain}
\newtheorem{mytheorem}{Theorem}

\newtheorem{mylemma}[mytheorem]{Lemma}

\newtheorem{myproblem}{Problem}
\theoremstyle{definition}

\usepackage{pifont}
\usepackage{graphics,graphicx,color}

\catcode`\@=11
\@ifpackageloaded{beamer}{
}{
}
\catcode`\@=12

\newcommand{\sympl}{\eta}

\newcommand{\setZ}{{\mathbb{Z}}}
\newcommand{\Ztwo}{\setZ^2}

\newcommand{\defeq}{\overset{\text{def}}{=}}

\newcommand{\Bx}{\boldsymbol{x}}

\newcommand{\Bn}{\boldsymbol{n}}
\newcommand{\HH}{{\mathcal{H}}}
\newcommand{\BH}{\boldsymbol{\HH}}

\newcommand{\sigmaN}{{\sigma^2}}

\newcommand{\FrameOp}{S}

\newcommand{\OrthoOp}{O}

\newcommand{\BHSpread}{\boldsymbol{\Sigma}}
\newcommand{\BHWeyl}{\boldsymbol{L}}
\newcommand{\BHScat}{\boldsymbol{C}}

\newcommand{\diag}{\text{diag}}

\newcommand{\Leb}[1]{\mathcal{L}_{#1}}

\newcommand{\Indexset}{{\mathcal{I}}}

\newcommand{\Amb}{{\mathbf{A}}}

\newcommand{\SINR}{\text{\small{\rm{SINR}}}}

\newcommand{\EX}[1]{{\mathbf{E}}\{#1\}}

\newcommand{\Real}[1]{{\text{Re}}\{#1\}}

\newcommand{\Shift}{{\boldsymbol{S}}}

\newcommand{\Ind}[2]{\chi_{#1}(#2)}

\newcommand{\sFourier}{\mathcal{F}_s}

\newcommand{\Reals}{\mathbb{R}}
\newcommand{\RealsPlus}{\mathbb{R}_{+}}
\newcommand{\Complexes}{\mathbb{C}}

\newcommand{\taumax}{{\tau_d}}

\begin{document}
\title{
  Pulse Shaping, Localization and the Approximate Eigenstructure of LTV Channels 
}
\author{Peter Jung\\
  Fraunhofer German-Sino Lab for Mobile Communications (MCI) and the Heinrich-Hertz Institute, Berlin\\[.1em]
  \small{jung@hhi.fraunhofer.de}}
\specialpapernotice{(Invited Paper)}
\maketitle
\begin{abstract}
   In this article we show the relation between the theory of pulse shaping 
   for WSSUS channels and the notion of approximate eigenstructure for time--varying 
   channels. We consider pulse shaping for a general signaling scheme, called Weyl--Heisenberg signaling,
   which includes OFDM with cyclic prefix and OFDM/OQAM.
   The pulse design problem in the view of optimal WSSUS--averaged $\SINR$ is an interplay between
   localization and ''orthogonality''. The localization problem itself can be
   expressed in terms of eigenvalues of localization operators and 
   is intimately connected to the concept of approximate eigenstructure of LTV 
   channel operators. In fact, on the $\Leb{2}$--level both are equivalent as we will show. 
   The concept of ''orthogonality'' in turn can be related to notion of tight frames. The right balance between these
   two sides is still an open problem. However, several statements on achievable values of
   certain localization measures and fundamental limits on $\SINR$ can already be made as will be shown 
   in the paper.
\end{abstract}

%
\section{Introduction}
Optimal signaling through linear time--varying (LTV) channels is a challenging
task for future communication systems. For a particular realization 
of the time--varying channel operator the transmitter and receiver
design which avoids crosstalk between different time--frequency slots
is related to ''eigen--signaling'' which simplifies much the
information theoretic treatment of communication in dispersive channels.
But it is well--known that
for an ensemble of channels which are dispersive in time and frequency (doubly--dispersive)
such a \emph{joint} separation of the 
subchannels can not be achieved because the eigen decompositions can 
differ from one to another channel realization.
Several approaches 
are proposed to describe ''eigen''--signaling in some approximate sense.
However, then a necessary prerequisite is the characterization of 
approximation errors.

A typical scenario, present for example
in wireless communication, is signaling through a random doubly--dispersive
channel, represented by the operator $\BH$. The received signal $r:\Reals\rightarrow\Complexes$ 
at time instant $t$ is then:
\begin{equation*}
   r(t)=(\BH s)(t)+n(t)
\end{equation*}
A preferred design of the transmit signal $s:\Reals\rightarrow\Complexes$ needs knowledge
on the true eigenstructure of the channel operator $\BH$. This would in principle
allow interference--free transmission and simple recovering algorithms of the information
from received signal $r$ degraded by the noise process $n$. 
However, for $\BH$ being a random operator, a random eigenstructure has to be 
expected in general such that a \emph{joint design} of the transmitter and the receiver
for an ensemble of channels has to be performed. Nevertheless,
interference can then not be avoided and
remains in the communication chain. For such interference scenarios 
bounds on the distortion of a particular selected 
signaling scheme are mandatory for reliable system design.

This problem has been considered, for example, already in \cite{kozek:eigenstructure} and  
\cite{matz:timefreq:characterization}. The investigation of the $\Leb{p}$--norm $E_p$ 
of the error $\BH s-\lambda r$ for $\lambda\in\Complexes$ enables us to improve and generalize recent results in this
direction. We will focus in this article on the results for $p=2$
which show the important relation to pulse shape optimization for 
Weyl--Heisenberg signaling 
\cite{kozek:thesis,jung:wssuspulseshaping,jung:isit06}. We will use this
generalized description of a multicarrier system and consider in this
framework the channel averaged ''signal to interference and noise ratio''.
To perform this average we will use the wide--sense stationary uncorrelated
scattering (WSSUS) channel model.

The paper is organized as follows. In Section \ref{sec:timevarying:channels} 
we will review the basics from time--frequency analysis, introduce the
spreading representation of doubly--dispersive channels and
consider the problem of controlling the approximation error $E_2$. 
In Section \ref{sec:signaling:and:pulsedesign} we will focus on the description of multicarrier transmission
and WSSUS pulse shape optimization. We demonstrate the relation to $E_2$.
In Section \ref{sec:loc:sup} we give 
several relations to the ''size'' (Lebesgue measure) $|U|$ of the support
$U$ in the time--frequency plane of the channel's spreading function. 
In the last part we will verify our framework with some numerical tests.


\section{Time--Varying Channels}
\label{sec:timevarying:channels}
\subsection{Time--Frequency Shifts}
The fundamental operations on a signal $\gamma:\Reals\rightarrow\Complexes$ caused by the 
time--varying nature of typical wireless channels are time--frequency shifts. Let us 
denote a displacement in time by $\mu_1$ and in frequency by $\mu_2$ by the operator
$\Shift_\mu$ defined as:
\begin{equation}
   (\Shift_{\mu} \gamma)(t):=e^{i2\pi\mu_2 t}\gamma(t-\mu_1)
   \label{eq:weyl:shift:shiftoperator}
\end{equation}
($i$ is the imaginary unit and $\mu=(\mu_1,\mu_2)$). These operators act isometrically on
the $\Leb{p}(\Reals)$ spaces\footnote{
  $\Leb{p}(\Reals^n)$ denotes Lebesgue spaces on $\Reals^n$ with usual
  norms $\lVert\cdot\rVert_p$ for $1\leq p\leq\infty$}, 
hence unitary on the Hilbert space $\Leb{2}(\Reals)$.
The abstract action of $\Shift_\mu$
onto a function $\gamma$ is observable for example as a correlation response
with another function $g$. In this way the (cross) ambiguity function:
\begin{equation}
   \Amb_{g\gamma}(\mu)\defeq\langle g,\Shift_\mu\gamma\rangle
   =\int_{\Reals}\bar{g}(x)\gamma(x-\mu_1)
   e^{i2\pi\mu_2x}dx
   \label{eq:tfanalysis:crossamb}
\end{equation}
can be defined as an observable quantity. 
Definition 
\eqref{eq:weyl:shift:shiftoperator} is not unique and corresponds to a
particular polarization in 
the Weyl--Heisenberg group (see \cite{folland:harmonics:phasespace,jung:approxeigen}). 

A fundamental operation in time--frequency analysis is the symplectic Fourier transform
$\sFourier$ defined for a function $f:\Reals^{2}\rightarrow\Complexes$ as
\begin{equation}
   (\sFourier f)(\mu)=\int_{\Reals^{2}} e^{-i2\pi\sympl(\nu,\mu)} f(\nu)d\nu
   \label{eq:tfanalysis:symplectic:fourier}
\end{equation}
where $\sympl(\nu,\mu):=\nu_1\mu_2-\nu_2\mu_1$ is called the symplectic form.
\eqref{eq:tfanalysis:symplectic:fourier} differs from ordinary Fourier 
transform only by coordinate/sign switching 
which quite intuitive because time--shifts $\mu_1$ cause oscillations in frequency where
frequency shifts $\mu_2$ cause them in time.
Then the (cross) Wigner distribution is the symplectic Fourier transform of the (cross) ambiguity function.

\subsection{Spreading Representation of Doubly--Dispersive Channels}

\newcommand{\OpClass}{\text{OP}}
Doubly--dispersive channels
are physically characterized 
by scattering in the time--varying environment. 
Typically one defines therefore the channel operator $\BH$
weakly in terms of the transmit and receive filters $\gamma$ and $g$
of a particular communication device as:
\begin{equation}
   \langle g,\BH \gamma\rangle=\int_{\Reals^{2}} \BHSpread_{\BH}(\mu)\Amb_{g\gamma}(\mu)d\mu  
   \label{eq:tfanalysis:linop:spreading}
\end{equation}
The function  $\BHSpread_{\BH}:\Reals^2\rightarrow\Complexes$ 
is then called the spreading function of the channel operator
$\BH$. Its symplectic Fourier transform, 
$\BHWeyl_{\BH}=\sFourier\BHSpread_{\BH}$, is called
the symbol of $\BH$. 
The practical assumption that the support $U\subset\Reals^2$ of $\BHSpread_{\BH}$ is compact 
renders $\BH$ to be a Hilbert--Schmidt operator whenever $\BHSpread_{\BH}$ is bounded and 
$|U|>0$ (single-- or non--dispersive channels have $|U|=0$). 
In this paper we will restrict the analysis to those channels to avoid the use of distributions.
To this end, let us call $\OpClass(U)$ as the set of channel operators $\BH$ having a spreading
function $\BHSpread_{\BH}$ with non--zero and finite support in $U$ where $0<|U|<\infty$.

\subsection{Approximate Eigenstructure}
It is a fundamental question, how well we can describe the action of the channel in
terms of its symbol. More generally speaking, how valid is an approximation  of the form:
\begin{equation}
   \BH\Shift_\mu\gamma\approx\lambda(\mu)\Shift_\mu g
\end{equation}
See for example \cite{durisi:wssus:capacity} for recent applications of this approximation.
For example, for $g$ and $\gamma$ being singular functions of $\BH$ (eigenfunctions
of $\BH\BH^*$ and $\BH^*\BH$) there is equality for a particular $\lambda(0)$ (the singular value). 
However, this is
not well suited for our formulation, because $\BH$ will be a random channel operator as for
example further described in Section \ref{subsec:wssus}. However, if we restrict the set of operators to be 
from class $\OpClass(U)$
we can seek for $g$ and $\gamma$ which are
independent of the channel realization and reasonable bound the 
approximation error $E_p=\lVert\BH\Shift_\mu\gamma-\lambda(\mu)\Shift_\mu g\rVert_p$.
For a single $\BH$ we will call this setup as \emph{approximate eigenstructure}\footnote{
  This approach is similar to the concept of a pseudospectrum.
  The set of 
  numbers $\lambda$ for which exists a normalized $\gamma$ such that
  $\lVert\BH\gamma-\lambda\gamma\rVert_2\leq\epsilon$ is called 
  the $\epsilon$--pseudospectrum of $\BH$.}  and
in this article we will consider $p=2$  because of its relation to pulse design
(considered in Section \ref{sec:signaling:and:pulsedesign}). For more general results see \cite{jung:approxeigen}.
Our problem is:
\begin{myproblem}
   Let be $\BH\in\OpClass(U)$ and $\lVert g\rVert_2=\lVert\gamma\rVert_2=1$ and $1\leq a\leq\infty$.
   Let $\delta\in\RealsPlus$ be {\bf independent} of $\BHSpread_{\BH}^{(\alpha)}$ 
   such that
   \begin{equation}
      \begin{split}
         E_2
         \leq \delta\cdot \lVert\BHSpread_{\BH}\rVert_a
      \end{split}
      \label{eq:approxeigen:epweyl}
   \end{equation}
   How small can we choose $\delta$ given $g$, $\gamma$, $U$ and $a$ ? 
   What can be said about $\inf_{g,\gamma}(\delta)$ given $U$ and $a$.
   \label{problem:approxeigen:epweyl}
\end{myproblem}
Note that this formulation comprises a 
\emph{joint} approximate eigenstructure for
all channels $\BH\in\OpClass(U)$ but with 
individual bounds  \mbox{$\delta\cdot \lVert\BHSpread_{\BH}^{(\alpha)}\rVert_a$}
for the approximation error.

First decompositions results in this field can be found 
already in the literature on pseudodifferential operators 
\cite{kohn:pdo:algebra,folland:harmonics:phasespace}.
More recent results with direct application to time--varying channels
were obtained by Kozek \cite{kozek:thesis}
and Matz \cite{matz:thesis} which resemble the notion of underspread
channels. They found results on $\delta$ for $\lambda=\BHWeyl_{\BH}$ and
$a=1$ which follow from the approximate product rule 
in terms of Weyl symbols. This estimate intimately scales with $|U|$ and
breaks down at a certain critical size. Channels 
below this critical size are called in their terminology underspread, 
otherwise overspread. However, the critical term in their estimate can be
dropped and the overall result can be improved to
$\delta^2\leq 2\lVert (1-\Amb_{g\gamma})\chi_U\rVert_\infty$ \cite{jung:approxeigen}. 
Similar estimates can be obtained for other weight functions instead of 
$\chi_U$ (the characteristic function of $U$). 
For this case ($a=1$) it is difficult to separate explicit 
relations on $|U|$. However, as will be shown, for $a>1$ this can be done.

With the choice $\lambda=\BHWeyl_{\BH}$ a formulation independent
of $g$ and $\gamma$ is achieved. An approach which is better suited for our application
instead is 
\begin{equation}
   \lambda(\mu)=\langle \Shift_\mu g,\BH,\Shift_\mu\gamma\rangle
   =\sFourier(\BHSpread_{\BH}\Amb_{g\gamma})(\mu)
\end{equation}
This choice is also known as the  orthogonal distortion (smoothing with the cross Wigner function)
and corresponds to the $E_2$ minimizer in \eqref{eq:approxeigen:epweyl}
for fixed $g$ and $\gamma$.
This approach was also already considered for $g=\gamma$ in \cite{matz:timefreq:characterization}.
One can prove the following Lemma \cite{jung:approxeigen}:
\begin{mylemma}
   Let $1<a\leq\infty$ and $1/a+1/b=1$ and 
   $\BHSpread_{\BH}\in\Leb{a}(\Reals^{2})$. If $|U|\leq 1$ it
   holds 
   \begin{equation}
      \frac{E_2}{\lVert\BHSpread_{\BH}\rVert_a}\leq
      \left(|U|-\int_{U}|\Amb_{g\gamma}(\mu)|^2d\mu\right)^{1/\max(b,2)}
      \label{eq:approxeigen:lemmaep1:1}
   \end{equation}   
   \label{lemma:approxeigen:lemmaep1}
\end{mylemma}
Similar bounds can be found for all $p$--norms with 
$2\leq p<\infty$ and more detailed results avoid also the constraint $|U|\leq1$.

\subsection{WSSUS Scattering}
\label{subsec:wssus}
A common statistical model for random doubly--dispersive channels is the WSSUS
channel model, originally introduced by Bello \cite{bello:wssus}. In this model
the spreading function $\BHSpread_{\BH}$ is described as a zero--mean
2D--random process uncorrelated at different arguments, which gives:
\begin{equation}
   \EX{|\langle g,\BH \gamma\rangle|^2}=\int_{\Reals^{2}} \BHScat(\mu)|\Amb_{g\gamma}(\mu)|^2d\mu  
   \label{eq:wssus:scattering}
\end{equation}
The (non--negative) function  $\BHScat:\Reals^2\rightarrow\RealsPlus$ is called the 
\emph{scattering function}. Let be  $\lVert\BHScat\rVert_1=1$. Then 
$\BHScat(\mu)$ is also the probability density for the occurrence
of a time--frequency shift $\mu$ in the channel. In time--frequency 
analysis the term $\langle \BHScat,|\Amb_{g\gamma}|^2\rangle$ in \eqref{eq:wssus:scattering} is sometimes
also called the modulation
2--norm \cite{feichtinger:modspaces} of $\gamma$ with respect to window $g$ and weight function $\sqrt{\BHScat}$.

\section{Signaling and Pulse Shaping}
\label{sec:signaling:and:pulsedesign}
In Lemma \ref{lemma:approxeigen:lemmaep1} and \eqref{eq:wssus:scattering} 
we have already shown certain localization measures characterizing
the signal distortion in time--varying channels, in particular for the WSSUS model.
In this section we will now establish a simple communication chain, including OFDM
and OQAM/OFDM, which use
$g$ and $\gamma$ as corresponding transmit and receiver filters. Finally, we aim at a
relation between $\{g,\gamma\}$, the WSSUS--averaged $\SINR$ and
the role of time--frequency localization which is relevant for $E_2$.

\subsection{Weyl--Heisenberg Signaling}
Conventional OFDM and pulse shaped OQAM/OFDM can be formulated jointly  within the concept of 
Weyl--Heisenberg signaling \cite{jung:thesis}.
To avoid cumbersome notation we will adopt a two--dimensional index notation
$n=(n_1,n_2)\in\setZ^2$ for time--frequency slots $n$, such that
the baseband transmit signal  is
\begin{equation}
   \begin{aligned}
      s(t)
      =\sum_{n\in\Indexset}x_n\gamma_n(t)
      =\sum_{n\in\Indexset}x_n(\Shift_{\Lambda n}\,\gamma)(t)
   \end{aligned}
   \label{equ:txsignal}
\end{equation}
where $\Lambda\Ztwo$ is a lattice ($\Lambda$ denotes its $2\times 2$ real generator matrix)
with $|\Lambda|:=|\det(\Lambda)|>0$.
The indices $n$  range over 
the doubly-countable set $\Indexset\subset\Ztwo$, referring to the data burst to be transmitted. 
In practice $\Lambda$ is often restricted to be diagonal, i.e. $\Lambda=\diag(T,F)$.
Other lattices are considered for example in \cite{strohmer:lofdm2}.
The efficiency (in complex symbols) of the signaling is $|\Lambda|^{-1}$.
The coefficients $x_n$ are the complex 
data symbols at time instant $n_1$ and subcarrier index $n_2$ 
(from now on $\,\bar{\cdot}$ always denotes complex conjugate and
$\cdot^*$ means conjugate transpose), where $\Bx=(\dots,x_n,\dots)^T$.
By introducing the elements 
$H_{m,n}:=\langle g_m,\BH\gamma_n\rangle$
of the channel matrix $H$
the multicarrier transmission can be formulated as the linear equation
\mbox{$\tilde{\Bx}=H\Bx+\tilde{\Bn}$},
where $\tilde{\Bn}=(\dots,\langle g_m,n\rangle,\dots)^T$ is the vector of the projected noise
having variance $\sigmaN:=\EX{|\langle g_m,n\rangle|^2}$ per component.
\subsection{Complex Signaling and OFDM}
\label{subsec:ofdm}
The classical OFDM system exploiting a cyclic prefix (cp-OFDM) is obtained by assuming a
lattice generated by $\Lambda=\diag(T,F)$ and 
$\gamma(t)\sim\Ind{[-T_{cp},T_u]}{t}$ which is (up to normalization)
the characteristic function of the interval $[-T_{cp},T_u]$.
$T_u$ usually denotes the duration of the useful part of the signal and $T_{cp}$ is the length of the cyclic prefix 
such that $T=T_u+T_{cp}$. 
The OFDM subcarrier spacing is $F=1/T_u$.
At the receiver the rectangular pulse $g(t)\sim\Ind{[0,T_u]}{t}$ is used which removes
the cyclic prefix. The efficiency  is given as
$|\Lambda|^{-1}=T_u/(T_u+T_{cp})<1$. It can be easily verified that 
$\Amb_{g\gamma}(\mu+\Lambda m)\sim\delta_{m,0}$ if $0\leq\mu_1\leq T_{cp}$ and $\mu_2=0$
(or see \cite{jung:ieeecom:timevariant} for the full formula).
Thus, orthogonality is preserved in time--invariant channels with maximum delay smaller than
$T_{cp}$, but at the cost of signal power (the redundancy is not used) and efficiency. 

\subsection{Real Signaling and OQAM}
\label{subsec:oqam}
For those schemes an inner product $\Real{\langle\cdot,\cdot\rangle}$ and  $\Lambda=\diag(T,F)$
having $|\Lambda|=1/2$
is considered, which
is realized by OQAM based modulation for OFDM (OQAM/OFDM) \cite{chang:oqam}. 
Before modulation the mapping
$x_n=i^n x^\text{R}_n$ has to be applied\footnote{We 
  use here the notation $i^n=i^{n_1+n_2}$. Furthermore other phase mappings are
  possible, like $i^{n_1+n_2+2n_1n_2}$.},
where $x^\text{R}_n\in\mathbb{R}$ is the real-valued information to transmit.
After demodulation
$\tilde{x}^\text{R}_m=\Real{i^{-m}\tilde{x}_m}$ is performed. Moreover, the pulses $(g,\gamma)$
have to be real. Thus, formally the transmission of the real information
vector $\Bx^\text{R}=(\dots,x^\text{R}_n,\dots)^T$ can be written as
$\tilde{\Bx}^\text{R}=H^\text{R}\Bx^\text{R}+\tilde{\Bn}^\text{R}$
where the real channel matrix elements are:
\begin{equation}
   H^\text{R}_{m,n}=\Real{i^{n-m}H_{m,n}}=\Real{i^{n-m}\langle g_m,\BH \gamma_n\rangle}
   \label{eq:hmn:realschemes}
\end{equation}
and ''real--part'' noise components are $\tilde{n}^\text{R}_m=\Real{i^{-m}\langle g_m,n\rangle}$.
There exists no such orthogonality property for OQAM based signaling
as for cp-OFDM in time--invariant channels, but 
biorthogonality of the form $\Re\{\langle g_m,\gamma_n\rangle\}\sim\delta_{m,n}$ 
can be achieved. It is known that the design of (bi--)orthogonal OQAM signaling
is implicitly related to (bi--) orthogonal Wilson bases \cite{boelckei:oqam} which is
in turn equivalent to the computation of (dual--) tight frames \cite{daubechies:simplewilson}.

While the system operates with real information at $\epsilon=2$ the effective efficiency is one, 
which has advantages in the view of pulse shaping as will explained later on. 

\subsection{A Lower Bound on the WSSUS Averaged $\SINR$}
\label{subsec:sinr}
In multicarrier transmission most commonly one--tap equalization per
time--frequency slot is considered, hence it is
naturally to require $\EX{|H_{m,m}|^2}$ to be maximal 
and the averaged interference power 
from all other lattice points to be minimal as possible. For real schemes $H_{m,m}$ has
to be replaced by $H^\text{R}_{m,n}$ from \eqref{eq:hmn:realschemes}.
For $\lVert\BHScat\rVert_1=1$, $\lVert g\rVert_2=\lVert\gamma\rVert_2=1$ and $\EX{\bar{x}_mx_n}=\delta_{m,n}$ we have shown
in \cite{jung:wssuspulseshaping} that the averaged 
''signal to interference and noise ratio'' is lower bounded by:
\begin{equation}
   \begin{aligned}
      \SINR(g,\gamma,\Lambda)\geq
      \frac{\langle\BHScat,|\Amb_{g\gamma}|^2\rangle}{\sigmaN+B_\gamma-\langle\BHScat,|\Amb_{g\gamma}|^2\rangle}
   \end{aligned}      
   \label{eq:sinr:bound}
\end{equation}
with equality for $\{\gamma_m\}$ forming a tight frame 
(more details and references on frame theory in \cite{jung:wssuspulseshaping}).
The constant $B_\gamma=\rho(\FrameOp_{\gamma,\Lambda})$ is the spectral radius
of the positive semidefinite operator $\FrameOp_{\gamma,\Lambda}$ defined as:
\begin{equation*}
   (\FrameOp_{\gamma,\Lambda} f)(t):=\sum_{\lambda\in\Lambda\Ztwo}
   \langle\Shift_\lambda\gamma,f\rangle(\Shift_\lambda\gamma)(t)
   \label{eq:frameop}  
\end{equation*}
The minimal $B_\gamma$  is $\max(|\Lambda|^{-1},1)$ which is achieved for tight frames 
in the case of $|\Lambda|\leq 1$ and for 
incomplete orthogonal bases in the case of $|\Lambda|>1$.
In addition to complex signaling as considered in Section \ref{subsec:ofdm}  
Equation \eqref{eq:sinr:bound} holds for real signaling (Section \ref{subsec:oqam}) as well if 
the spreading function and the noise are circular--symmetric processes \cite{jung:wssuspulseshaping}.

\subsection{$\SINR$ Optimization and the Pulse Shaping Problem}
\label{subsec:pulseshaping:problem}
It follows that maximizing the bound in \eqref{eq:sinr:bound} is an interplay between two
criteria: \emph{localization} (maximizing  $\langle\BHScat,|\Amb_{g\gamma}|^2\rangle$) and 
\emph{''orthogonality''}, respectively ''tightness'' (minimizing $B_\gamma$). 
Orthogonalization is only possible for $|\Lambda|\geq1$. 
For $|\Lambda|\leq1$  a tight frame has to be constructed. However, 
with the notion of the adjoint lattice $\Lambda^\circ:=|\Lambda|^{-1}\Lambda$
both cases are dual in the sense of Ron and Shen \cite{ronshen:duality}, which means that
we can limit ourself to the case $|\Lambda|\leq 1$.

One possible method \cite{strohmer:lofdm2,jung:wssuspulseshaping} to achieve 
a minimal $B_\gamma$ in a $\Leb{2}$--optimal sense is to apply
$\OrthoOp:=\left(|\Lambda|\cdot\FrameOp_{\gamma,\Lambda}\right)^{-\frac{1}{2}}$ on the result of a localization procedure\footnote{
  $\FrameOp_{\gamma,\Lambda}^{-\frac{1}{2}}$ exists if $\{\gamma_n\}_{n\in\setZ^2}$ is a frame.
} to obtain $\gamma^\circ=\OrthoOp\gamma$ (for $|\Lambda|>1$ a similar procedure follows from
replacing $\Lambda$ by $\Lambda^\circ$). For example,
in \cite{janssen:equivalencefab} it was shown that 
IOTA (Isotropic Orthogonal Transform Algorithm, $\Lambda=\diag(T,F)$ with $|\Lambda|=1/2$) 
as proposed in \cite{lefloch:cofdm} is equivalent to $\OrthoOp$ 
applied on a Gaussian. In combination with OQAM modulation as introduced in section \ref{subsec:oqam} this
is also an orthogonal signaling (in the absence of a channel).
The advantage of this approach is that at $|\Lambda|=1/2$ the operation
$\OrthoOp$ is quite stable. In contrast: optimal
efficiency with orthogonal (or bi--orthogonal)
complex signaling (this implies $|\Lambda|=1$) is always affected by the Balian--Low
Theorem (BLT) (see for example \cite{grochenig:gaborbook}), i.e. 
$\OrthoOp$ is unbounded if it exists.

Estimates on the localization penalty of the operation $\OrthoOp$ on the $\SINR$ 
are in general related to mapping properties of $\OrthoOp$ on 
modulation spaces \cite{feichtinger:modspaces}. Beside the general problem of 
invertibility of the operator in \eqref{eq:frameop} (for recent results
see here for example \cite{groechenig:wiener:twisted} or \cite{grochenig:gaborbook})
there remains the open question on optimal constants which are necessary to draw
conclusions on the $\SINR$.

\section{Localization and Support Bounds}
\label{sec:loc:sup}
\subsection{Pulse Shaping and Approximate Eigenstructure}
The detailed knowledge of the scattering function of an statistical ensemble of
channels can improve the overall performance of a communication device \cite{jung:wssuspulseshaping}.
However, also this statistical parameters change over time in realistic applications. 
A more reliable assumption is the support knowledge only, i.e. a scattering function which
is the normalized indicator function $\BHScat=\chi_U/|U|$ of $U$. We then see that approximate
eigenstructure in the formulation given in Lemma \ref{lemma:approxeigen:lemmaep1}  
and the localization aspect (the term
$\langle \BHScat,|\Amb_{g\gamma}|^2\rangle$) 
of pulse shaping \eqref{eq:sinr:bound} are the same problems.

\subsection{Localization Operators}
Since $|\Amb_{g\gamma}|^2$ is quadratic in $\gamma$ we can rewrite 
$\langle |\Amb_{g\gamma}|^2 ,\BHScat\rangle=\langle \gamma,L_{\BHScat,g}\gamma\rangle$ where this quadratic
form defines (weakly) an operator $L_{\BHScat,g}$. 
Such operators are also called \emph{localization operators} \cite{daubechiez:tflocalization:geophase}.
$L_{\BHScat,g}$ is compact which follows from our assumptions on $\BHScat$, such that
\begin{equation}
   \max_{\gamma}\langle |\Amb_{g\gamma}|^2,\BHScat\rangle=\lambda_{\max}(L_{\BHScat,g})
   \label{eq:approxeigen:gainmax}
\end{equation}
This means that for fixed $g$ the optimal transmit pulse $\gamma$ is given by an eigenfunction
corresponding to the maximal eigenvalue of $L_{\BHScat,g}$. The latter can also be reversed,
i.e. for fixed $\gamma$ the optimal receive filter $g$ is determined.
\begin{equation}
   \max_{g}\langle |\Amb_{g\gamma}|^2,\BHScat\rangle=\lambda_{\max}(L_{\tilde{\BHScat},\gamma})
   \label{eq:approxeigen:gainmax:g}
\end{equation}
where $\tilde{\BHScat}(\mu)=\BHScat(-\mu)$.
The eigenvalues and eigenfunctions of Gaussian ($g$ is set to be a Gaussian)
localization operators on the disc ($U$ is a disc) or more generally
with $\BHScat$ having elliptical symmetry are known to be Hermite 
functions \cite{daubechiez:tflocalization:geophase}. Kozek \cite{kozek:thesis} found that
for elliptical symmetry also the joint optimization ($g$ \emph{and} $\gamma$) results in Hermite 
functions\footnote{Kozek considered $g=\gamma$. However one can show that
  for elliptical symmetry around the origin the optimum has also this property.}.
Explicitely known is the joint optimum  for $\BHScat$ being 
Gaussian \cite{jung:isit06}. 
In \eqref{eq:approxeigen:gainmax} there is an invariance with respect to (affine) canonical transformations.
See for example \cite{folland:harmonics:phasespace}
for a review on the metaplectic representation. Essentially, this means that we can 
translate, rotate and shear $\BHScat$ into some 
prototype shape being canonical equivalent 
but with further symmetries.

Focusing now on localization operators $L_{\BHScat,\gamma}$ with 
$\BHScat:=\chi_U/|U|$ and let be $\lambda_{\max}:=\lambda_{\max}(L_{\BHScat,\gamma})$.
One can show the following result:
\newcommand{\erf}{\text{erf}}
\begin{mylemma}
   Let $U$ canonical equivalent to square. Then: 
   \begin{equation}
      \frac{\erf(\sqrt{\pi|U|/4})^4}{|U|^2}\leq\sup_{\lVert\gamma\rVert_2=1}\lambda_{\max}
      < \min(e^{-\frac{|U|}{e}},|U|^{-1})
      \label{eq:lemma:locbounds:rectangular}
   \end{equation}
   The upper bound is general and holds for any shape.
   \label{lemma:locbounds:rectangular}
\end{mylemma}
The proof can be found in \cite{jung:approxeigen}. The upper bound can also be found
in \cite{jung:isit06}.
The lower bound can be calculated for other shapes (like a disc) as well. As a result
for Problem \ref{problem:approxeigen:epweyl} 
it follows from Lemma \ref{lemma:approxeigen:lemmaep1} that:
\begin{equation}
   E_2\leq \lVert\BHSpread_{\BH}\rVert_a\cdot\left(|U|-\text{erf}(\sqrt{\pi|U|/4})^2\right)^{1/\max(b,2)}
\end{equation}
\subsection{Bounds for $\SINR$}
On the other hand, from the previous results we are also able to obtain for $|\Lambda|\leq1$ an upper 
bound on  $\SINR$ as follows.
Assume that $\{\gamma_n\}_{n\in\setZ^2}$ establishes a frame. Then we can compute
$\gamma^\circ=\OrthoOp\gamma$ to obtain a 
tight frame $\{\gamma^\circ_n\}_{n\in\setZ^2}$ with 
$B_{\gamma^\circ}=|\Lambda|^{-1}$. For a tight frame 
holds equality in \eqref{eq:sinr:bound}. It remains to find the optimal $g$ which can
be found for example as the maximizing (generalized) eigenvalue as shown in \cite{jung:wssuspulseshaping}
(see also Section \ref{subsec:approxeigen:localg}). Let us call $\SINR^*:=\max_g\SINR(g,\gamma^\circ,\Lambda)$.
The upper bound 
in \eqref{eq:lemma:locbounds:rectangular} of Lemma \ref{lemma:locbounds:rectangular} gives
then for $|U|\leq e$ that:
\begin{equation}
   \SINR^*\leq
   \left((\sigmaN+|\Lambda|^{-1})e^{\frac{|U|}{e}}-1\right)^{-1}
   \label{eq:approxeigen:upper:sinr}
\end{equation}
Effectively, the upper bound \eqref{eq:approxeigen:upper:sinr} represents the
assumption that we would be able to retain with $\gamma\rightarrow\gamma^\circ$ the 
localization upper bounded
by the rhs of \eqref{eq:lemma:locbounds:rectangular}, thus ignoring the loss due to
the BLT. This approach is valid to obtain an upper bound.

As already discussed at the end of Section \ref{subsec:pulseshaping:problem},
in general no quantitative estimates exists on the localization loss, i.e.
it is unknown to what degree \eqref{eq:approxeigen:upper:sinr} is achievable.
Assuming negligible localization loss (''no BLT'') the lhs of \eqref{eq:lemma:locbounds:rectangular}
gives us also that:
\begin{equation}
   \SINR^*\overset{\text{no BLT}}{\geq}
   \left((\sigmaN+|\Lambda|^{-1})\frac{|U|}{\erf(\sqrt{\pi|U|/4})^4} -1\right)^{-1}
   \label{eq:approxeigen:lower:sinr}
\end{equation}
For uncritical $\Lambda$ (or $\Lambda^\circ$ respectively) the mapping
$\OrthoOp$ can be well conditioned (which still depends on $\gamma$) such that
\eqref{eq:approxeigen:lower:sinr} is an estimate of achievable values of $\SINR^*$.
Our numerical results, shown in Section \ref{subsec:approxeigen:numver}, support these assumptions
for $|\Lambda|=1/2$.

\section{Localization Algorithms and Numerical Verification}
\label{sec:localalg:num}
A well--known approach is to adapt the lattice $\Lambda$ and the time-/frequency spreads $\sigma_t$ and
$\sigma_f$ of the pulses to
the scattering function $\BHScat$ of an ensemble of WSSUS channels where the device should operate.
In practice various environments have to be classified by their
spread in delay ($\BHScat_t$) and mobility (spread $\BHScat_f$ in two--sided Doppler frequencies)
such that a potential mobile device will support a group of different modes $\{g,\gamma,\Lambda\}$.
For $\Lambda=\diag(T,F)$ this means to fulfill
essentially $T/F\approx\sigma_t/\sigma_f\approx\BHScat_t/\BHScat_f$ which is known as
''pulse and grid scaling'' \cite{kozek:thesis,liu:orthogonalstf,jung:wssuspulseshaping}. However, up to scaling (and displacing) this approach does not further
fix $g$ and $\gamma$. From time--frequency uncertainty reasons a Gaussian or 
the ''tighten'' Gaussian (the IOTA approach) can be used for example. 
More detailed properties of the scattering function can be exploited in pulse design
with the following methods directly based in the theory presented so far.

\subsection{''Mountain Climbing''}
\label{subsec:approxeigen:localg}
The general $\SINR$ problem is known to be a quasi--convex, 
convex--constrained maximization problem \cite{jung:wssuspulseshaping}, that
is a \emph{global optimization problem}. 
The same holds, separately, for the localization problem: 
\begin{equation}   
   \sup_{g}\lambda_{\max}(L_{\BHScat,g})=
   \sup_{\gamma}\lambda_{\max}(L_{\tilde{\BHScat},\gamma})
   \label{eq:approxeigen:gainmax:g}
\end{equation}
as well, which can be formulated as a convex, convex--constrained maximization problem.
But, a so called ''mountain climbing'' algorithm can be used to perform a local optimization
(called ''Gain optimization'' in \cite{jung:wssuspulseshaping}). Essentially
this means to start with an appropriate $g^{(0)}$ (for example a Gaussian). In $n$th iteration step 
$\gamma^{(n)}$ is calculated as the maximizing eigenfunction
of $L_{\BHScat,g^{(n-1)}}$. From $\gamma^{(n)}$ the operator 
$L_{\tilde{\BHScat},\gamma^{(n)}}$ is constructed to compute $g^{(n)}$
as its maximizing eigenfunction. Similarly, the $\SINR$ can also be optimized directly
with the notion of generalized eigenvalues (more details in \cite{jung:wssuspulseshaping}).

\subsection{A Solvable Lower Bound}
\label{subsec:approxeigen:lower}
In contrast to the iteration method presented in the last section, it is possible 
to solve a lower bound of localization problem in one step. A simple convexity 
argument shows that:
\begin{equation}
   \langle \BHScat,|\Amb_{g\gamma}|^2\rangle\geq |\langle g,Q\gamma\rangle|^2
   \label{eq:approxeigen:lower:gain}
\end{equation}
where $Q$ is an operator with spreading function $\BHScat$. The optimum of the rhs
of \eqref{eq:approxeigen:lower:gain} over $g$ and $\gamma$ are the 
maximizing eigenfunctions of $QQ^*$ and $Q^*Q$ (on the level of matrices
computable by the SVD).

\subsection{Numerical Verification}
\label{subsec:approxeigen:numver}
In the following we will compare the performance on localization and $\SINR$
of the different methods presented above. We consider a pulse shaped 
OQAM/OFDM system as presented in \ref{subsec:oqam}, i.e.  $\Lambda=\diag(T,F)$ with
$|\Lambda|=1/2$. The scattering function is $\chi_U/|U|$ for $U=[0,\taumax]\times[-B_D,B_D]$
and $|U|\approx 0.29$ which is a highly interference dominated scenario (but still 
underspread). We have varied the ratio between time and frequency dispersion
(different $\taumax/(2B_D)$) and $\Lambda$ is always adapted according to the 
grid scaling rule (as explained above).
All computations are performed on a discrete time system of length $L=512$ (for
more details see \cite{jung:wssuspulseshaping}) and bandwidth $W$, such that 
the equivalent discrete variables are $\taumax^{(D)}:=\taumax\cdot W$ and $B_D^{(D)}:=B_D/W\cdot L$.
While keeping constant $(\taumax^{(D)}+1)(2\cdot B_D^{(D)}+1)=150$ the discrete 
ratio $R:=(\tau^{(D)}_d+1)/(2B_D^{(D)}+1)$ has been varied according to the following 
table:
\begin{center}
   \begin{tabular}{r|c|c|c|c|c|c|c}
      $\tau_d^{(D)}$ &0  &    1 &    5 &    9 &  29 &  49 &  149 \\\hline
      $B_D^{(D)}$    &74 &   37 &   12 &    7 &   2 &   1 &    0 \\     
   \end{tabular}
\end{center}
In summary we will show the following methods:
properly scaled Gaussians ({\bf Gauss}),
$\{g^\circ,\gamma^\circ\}$ from the method of
Section \ref{subsec:pulseshaping:problem} with properly scaled Gaussians as input ({\bf IOTA}),
maximizers of the 
rhs of  \eqref{eq:approxeigen:lower:gain} ({\bf SVD}),
algorithm of Section \ref{subsec:approxeigen:localg} ({\bf localg}).
$\{g^\circ,\gamma^\circ\}$ from the method of Section \ref{subsec:pulseshaping:problem} 
with ''localg'' as input ({\bf localg--tight}) and
the iterated  $\SINR$ algorithm explained due to limited
space only in \cite{jung:wssuspulseshaping} ({\bf sinralg}).
The last algorithm is in principle equivalent to
the eigenvalue optimization in ''localg'' for particular noise level (here $\sigma^2=-20$dB), 
however on the level of  generalized eigenvalues. 

We will compare them to the theoretical estimates:
the rhs in \eqref{eq:approxeigen:upper:sinr} ({\bf upper}) and
the rhs of \eqref{eq:approxeigen:lower:sinr} under the assumption that 
there will be no localization
loss due to the BLT ({\bf lower}).
\begin{figure}[h]
   \includegraphics[width=\linewidth]{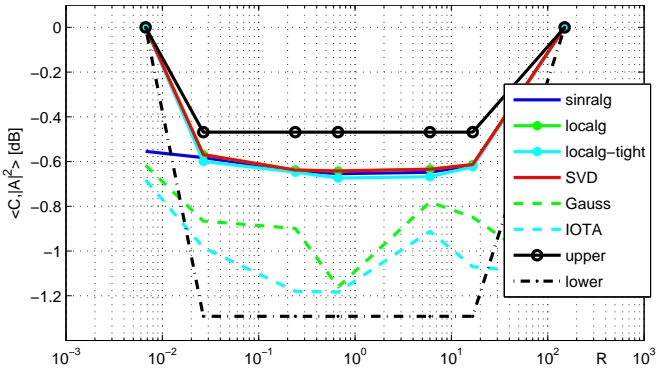}
   \caption{{\it Localization $\langle\BHScat,|\Amb_{g\gamma}|^2\rangle$ for $|U|\approx0.29$,
       lattice $|\Lambda|=1/2$ and OQAM signaling. $R$ has been varied.}}
   \label{fig:wssus:cmplx:gain:flat}
\end{figure}
\begin{figure}[h]
   \includegraphics[width=\linewidth]{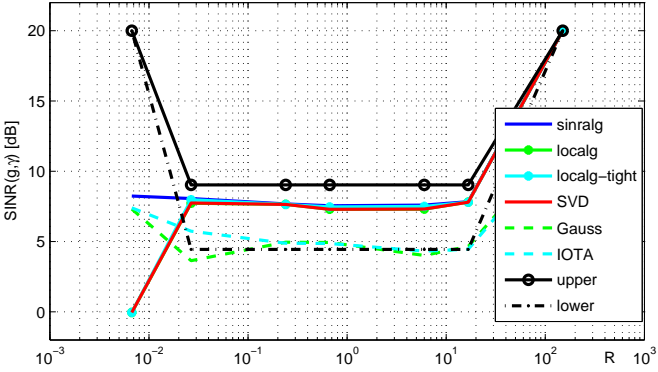}
   \caption{{\it $\SINR$ for  $|U|\approx0.29$ at noise power $\sigma^2=-20$dB, 
       lattice $|\Lambda|=1/2$ and OQAM signaling}. $R$ has been varied.}
   \label{fig:wssus:cmplx:sinr:flat}
\end{figure}

Fig.\ref{fig:wssus:cmplx:gain:flat} shows the corresponding  localization
\eqref{eq:wssus:scattering} 
and Fig.\ref{fig:wssus:cmplx:sinr:flat}
the $\SINR$.
It can be seen that the bounds are suited to describe the 
averaged performance of a multicarrier system in doubly--dispersive channels
based only the support parameter $|U|$. It is 
interesting that also lower bound gives a quite useful estimate on
the performance which is due to the uncritical condition number
of $\OrthoOp$ at lattice density $|\Lambda|^{-1}=2$. For OFDM systems
operating at ''more critical'' lattices much more loss to the BLT has 
to be expected.
The inner values of $R$ correspond to ''full'' doubly--dispersive channels where
no common eigenstructure can exists such that each decomposition is always 
an approximation.
The boundary values in turn 
are single--dispersive channels having a joint exact eigenstructure, 
i.e. it is possible to completely suppress interference.

\section{Conclusions}
We have introduced the theory of pulse shaping with focus on the 
WSSUS--averaged $\SINR$ and considered the question of approximate eigenstructure 
of time--varying channels with compactly supported spreading. 
With increasing demand on bandwidth and efficiency the understanding  
of the fundamental limits in both directions will be important
for future wireless communication systems.
For Weyl--Heisenberg signaling, as a general description of 
OFDM and OFDM/OQAM communication systems,
we found that both are on the level of localization  equivalent. 
We have shown that simple localization bounds can be used to obtain general
estimates on the eigenpair approximation behavior $E_2$ and
the $\SINR$ itself. With the latter, for example, we are able to show fundamental limits on achievable
performance based on simple statistical properties of the time--varying environment.

%

\bibliographystyle{IEEEtran}
\bibliography{references}
\end{document}